\documentclass[12pt,a4paper]{article}
\usepackage{mathptmx,graphicx}
\usepackage[top=1in, bottom=1in, left=1in, right=1in]{geometry}
\usepackage[round]{natbib}
\usepackage{afterpage}
\usepackage{xcolor}

\usepackage{hyperref}

\begin{document}

\title{\textsf{AI, Opacity, and Personal Autonomy}}
\author{Bram Vaassen}

\maketitle

\begin{quote}
\begin{center}

\textup{Penultimate draft, please cite published version}\\

\textup{Forthcoming in \emph{Philosophy and Technology}}: \href{https://link.springer.com/article/10.1007/s13347-022-00577-5}{10.1007/s13347-022-00577-5}\\

\end{center}
\end{quote}

\vspace{15pt}

\begin{abstract}
    \noindent Advancements in machine learning have fuelled the popularity of using AI decision algorithms in procedures such as bail hearings \citep{Feller2016}, medical diagnoses \citep{Rajkomar2018, Esteva2019} and recruitment \citep{Heilweil2019, Van2019}. Academic articles \citep{Floridi2018}, policy texts \citep{AIHLEG2019}, and popularizing books \citep{ONeill2016, Eubanks2018} alike warn that such algorithms tend to be \emph{opaque}: they do not provide explanations for their outcomes. Building on a causal account of transparency and opacity as well as recent work on the value of causal explanation \citep{Lombrozo2011, Hitchcock2012}, I formulate a moral concern for opaque algorithms that is yet to receive a systematic treatment in the literature: when such algorithms are used in life-changing decisions, they can obstruct us from effectively shaping our lives according to our goals and preferences, thus undermining our autonomy. I argue that this concern deserves closer attention as it furnishes the call for transparency in algorithmic decision-making with both new tools and new challenges.\\

\vspace{10pt}

\noindent \textbf{Keywords}: Explainable AI (XAI); Transparency; Opacity; Machine Learning; Autonomy

\end{abstract}

\pagebreak

\section{Introduction}
Artificial intelligence provides an increasingly popular way of streamlining important and sometimes life-changing decision procedures. Bail hearings \citep[e.g.][]{Feller2016}, medical diagnoses \citep[e.g.][]{Rajkomar2018, Esteva2019}, and recruitment procedures \citep[e.g.][]{Heilweil2019, Van2019} often rely on such automatized components. Their growing popularity is at least in part due to technical improvements in the design of machine learning algorithms, which drastically increased the amount of data that can be processed, as well as the rate at which it is processed. Crucially, such deep learning algorithms do not only use pre-programmed rules to provide outcomes. They also hone their procedures through numerous trial-and-error runs. Many of these AI decision algorithms are \emph{opaque} even when they are reliable: they might deliver the right results, but they do not provide users or affected parties any insight as to how they came to produce those results.\footnote{I'll use `users' in a broad sense here, to cover both those who employ and those who develop the algorithms. 
}

Many authors treat  opacity as one of the central threats posed by our increasing reliance on AI. Calls for transparency in automatized decision-making are commonplace in policy texts \citep[e.g.][]{AIHLEG2019, ICO2021}, academic literature \citep[see][]{Floridi2018}, and popularizing books \citep[e.g.][]{ONeill2016}, and are motivated in a variety of ways. I argue that there is a distinct concern with opaque algorithms that has not received a systematic treatment in the literature:  opacity can undermine the users' autonomy by hiding salient pathways of affecting the algorithm's outcomes. Importantly, even sufficiently reliable and fair\footnote{In what follows, I will use `fair' to mean `not biased in a problematic way' and assume that this notion of fairness is sufficiently well-understood. This terminological shortcut serves to avoid difficult questions such as what counts as (problematic) bias and whether or not opaque decision algorithms can be `fair' in broader sense.} algorithms can pose such a threat to autonomy. Consequently, this threat can remain even if other worries about opaque algorithms are successfully resolved and addressing this threat will require a certain degree of transparency in decision-making.

The text is structured as follows. First, I discuss previous motivations of the call for transparency, as well as some criticism on transparency demands (Section \ref{Call}). I then provide a causal account of opacity and transparency, and argue that it reveals how  opacity can undermine our autonomy. (Section \ref{Transparency}). Before concluding, I discuss some practical consequences of opacity's threat to our autonomy (Section \ref{Practice}). In particular, I discuss the impact of viewing transparency issues through the lens of our autonomy on (i) technological obstacles to delivering the right degree of transparency, (ii) prospects for legally entrenching adequate transparency demands, and (iii) weighing the advantages and drawbacks of transparent decision-making.

\section{The call for transparency}\label{Call}
Many participants in the explainable AI (XAI) debates call for transparency, explainability, explicability, or simply for less opacity. The general idea is that users of AI algorithms are entitled to explanations of their outcomes.\footnote{Some authors provide more fine-grained notions in this area as well, such as `traceability' and `simulatability' \cite[e.g.][]{Lipton2018}. Although I am convinced that the causal account I will present in Section \ref{Transparency} can capture most of these notions as well, I cannot argue for it here and the arguments presented below go through regardless. Our goal in this text is not to capture all the nuances in XAI terminology, but rather to pick out a moral concern with decisions that are not backed by explanations.} Quite often, AI algorithms are opaque in the sense that such explanations are not available to all stake-holders. This opacity can have different sources. Sometimes institutions or corporations fail to communicate when they rely on AI systems or on how these systems work.  Alternatively, those who are entitled to transparency might lack the expert knowledge to understand the explanations at hand. Finally, due to the increasing reliance on machine learning, even the most expertly trained humans might fail to grasp the algorithm in full detail.\footnote{See \citet{Burrell2016}, \citet{Weller2019},  and \citet{Walmsley2020} for more on these distinctions in opacity.} 
Our primary concern here is not what makes AI algorithms opaque, but rather why we should want them to be transparent.




 Transparency is treated as an important requirement for ethical AI implementations. In a survey of the literature on sustainable AI, \citet{Floridi2018} report that ``in all [texts surveyed], reference is made to the need to \emph{understand} and \emph{hold to account} the decision-making processes of AI.'' (p.~700, emphasis in original).\footnote{I am assuming here that understanding a decision-making process requires a certain amount of transparency, but it is important to note that there certainly isn't a one-to-one correspondence between them. More on this later.} The interest in transparency is by no means merely academic. It shows up in both policy texts and popularizing texts as well. For example, the EU guidelines for trustworthy AI specify that at a crucial component of the `transparency' they demand is to ``require[] that the decisions made by an AI system can be understood and traced by human beings.'' \citep[p.~18]{AIHLEG2019} and the recent European Commission's AI act proposes that ``a certain degree of transparency should be required for high-risk AI systems'' \citep[p.~30]{EUAI}. Finally, \citet[p.~31]{ONeill2016} lists opacity, damage and scale as the three essential features of algorithms that qualify as `weapons of math destruction'. There appears to be broad agreement that  transparency and opacity carry substantial moral weight.


This idea certainly has intuitive appeal. Receiving decisions without explanations can be frustrating and scary \citep[cf.][p.~29]{ONeill2016}. Empirical research confirms that our trust in decisions, actions and outcomes increase when we are provided with a plausible explanation \citep[e.g.][]{Herlocker2000, Symeonidis2009, Holzinger2020},\footnote{Although there are definitely other factors in play as well. See for example \cite{Krugel2022} and \cite{Jauernig2022}.} and earlier research in psychology suggests that there is a distinct pleasure associated with grasping explanations \citep{Gopnik1998}. Transparency comes with certain practical advantages as well. Our ability to assess the reliability and fairness of algorithms improves when we grasp the explanations for their outcomes \citep[]{Kim2016, Gkatzia2016, Biran2017a, Doshi2017} and improving an algorithm is probably easier when one knows how it works \citep[cf.][p.~18]{AIHLEG2019}.\footnote{Though see \citet[Sect. 3.1]{Weller2019} for examples suggesting that transparency can lead us to overascribe reliability.} By contrast, opacity often compounds the negative impact of inadequate outcomes. For example, it is harder to question the result of a process if one cannot point the finger at where it went wrong. Consequently, it is harder to negotiate the outcomes of opaque algorithms; their decisions are more likely to go unchallenged when mistaken and those responsible are less likely to be held to account \citep[cf.][]{ONeill2016, Wachter2017a, Floridi2018, Walmsley2020}.

Even so, the call for transparency is not without its critics. \citet{Zerilli2019} maintain that non-AI decision algorithms, such as humans or bureaucratic systems, can be equally complex and mysterious in their actual implementation. We rely on judges and committees to make decisions without having any deeper understanding of how their brain functions.\footnote{\citet{Lipton2018} and \citet{Cappelen2021} make similar points in passing.} \citet{London2019} emphasizes that common medical interventions often involve mechanisms that are not fully understood by patients, practitioners or researchers. By demanding full  transparency of AI decision algorithms, we seem to be holding them to a higher standard than we apply to non-AI decision makers or mechanisms. Reports explaining decisions on job applications never specify the neural details of the committee members, but provide us with very coarse-grained information: `Candidate A's references are unconvincing', `Candidate B lacks the relevant work experience', etc. Such explanations are typically located at the \emph{intentional} level: they concern the beliefs committee members have and how they integrate with their goal of getting the right candidate. An adequate standard of transparency for AI systems would thus similarly demand only restricted access to the workings of the system.\footnote{Though see \citet{Gunther2021}, who argue that a double standard would in fact be justified.}


Exactly how the demand for transparency should be restricted is a tricky question. It probably will not do to simply demand transparency at the intentional level, as many maintain that current artificially implemented algorithms do not have such a level \citep[cf.][]{Penrose1989, Dreyfus1992, Wakefield2003}. In this article, I identify a distinct moral concern with opaque decision algorithms. Even in cases where an opaque algorithm fulfills its task excellently, and its results are treated as negotiable, its opacity can threaten the autonomy of the parties affected by its outcomes. Getting clear on opacity's threat to autonomy will not only provide extra motivation for demanding transparency, it will also help to determine an adequate standard for transparency. Given that such a threat to our autonomy can occur regardless of whether the decision algorithm is executed by an AI system or by a human, motivating the demand for transparency with its relevance to autonomy would not require a double standard of the kind \citet{Zerilli2019} worry about.


\section{Transparency and autonomy}\label{Transparency}
In order to get clear on the tight relation between transparency and the autonomy of the affected parties, it will help to be more precise on what opacity and transparency actually consist of. I take the causal explanation literature to provide a natural account of transparency and opacity, as well as an intuitive explanation of their relation to autonomy. Let us start with the account of transparency and opacity.

\subsection{The causal account}

Champions of transparency maintain that users are entitled to an explanation of algorithm outcomes. When an algorithm decides that our job application does not make the cut, we are entitled to knowing why it didn't. A standard way of explaining why events occur is by referring to their causes \citep[cf.][]{Lewis1986, Woodward2003}. The drop in oil prices can be explained by the decreasing demand, because the decrease in demand caused the drop in prices. Similarly, my not getting the job can be explained by my lack of convincing references, if their poor quality caused my application to be rejected. According to the resulting picture, we can characterize opacity and transparency as follows:\footnote{Recently, \citet{Paez2019} argued that causal explanation is ill-suited to capture understanding of AI algorithm outcomes. I cannot address his concerns here, but see \citet{Erasmus2020, Erasmus2022} for a response. See also \citet{Frigg2009} for arguments in favour of non-exceptionalism about computer-run algorithms and simulations.}

\begin{quote}
\begin{description}

\item[\textsc{Opacity}] An AI system is  opaque to A to the extent that A is not in a position to grasp the causal explanations of its outcomes.

\item[\textsc{Transparency}] An AI system is  transparent to A to the extent that A is in a position to grasp the causal explanations of its outcomes.

\end{description}
\end{quote}
These characterizations need some unpacking. First and foremost, we need to know what a causal explanations are, and what it means to grasp them. Most off-the-shelf accounts from the philosophical literature will do for our purpose, but we will take a broadly interventionist approach as a starting point here. According to such accounts, $X$ causally explains $Y$ if and only if bringing about changes in $X$ whilst leaving other causes unmeddled with correlates with changes in $Y$, or with the chance of $Y$ occurring \citep[e.g.][p.~59]{Woodward2003}.\footnote{\citet{Wachter2017a} also focus on the importance of counterfactuals for transparency, but eschew overtly causal interpretations of these counterfactuals.} For example, if changing my references whilst leaving all else in my application fixed correlates with my success in getting the job, then my references causally explain my not getting the job. Simplifying quite a bit, we can say that to grasp a causal explanation of $X$ is to know what caused $X$ and to understand how changes in those causes correlate with changes in $X$.\footnote{See \citet{Strevens2013} for a more detailed account of grasping causal explanations. For our purposes, it is important that understanding how $X$ and $Y$ correlate does not require full understanding of the mechanism that makes them correlate. I can understand that changes in the position of the volume switch are proportional to changes in the volume output, without understanding the fine details of the relevant mechanism connecting the switch to the volume output.}

Second, it is worth emphasizing that such accounts of causal explanation are extremely non-committal about the physical or technical implementation of such counterfactual patterns. For example, as long the right pattern of counterfactual dependence holds between the quality of reference letters and the probability of getting the job, it does not matter how the importance of good reference letters is encoded in the system. In fact, it might very well be that there is no explicit mention of reference letters and what makes them convincing in the code of the algorithm. For many current AI algorithms, it is exceedingly likely that such information is encoded implicitly rather than explicitly \citep[cf.][p.~8--10]{Burrell2016}. According to the broadly counterfactual accounts of causation, being merely implicitly encoded is not an obstacle for being causally effective. As we shall see in Section \ref{Practice}, this feature will make such accounts quite suitable for providing explanations of AI behaviour at the right level.




A third point worth elaborating on is that opacity and transparency are matters of degree. One might understand how some outcomes of an AI system are produced, but not how others are. Similarly, one might know some of the causal factors contributing to outcome $X$ without knowing all of them. Crucially, not all causal explanations will help us \emph{understand} the outcome, nor will all causal explanations be of interest to us. For example, an overly inclusive causal explanation might confuse us, and thus provide us with no more grasp of how the outcome was produced. On the other end of the spectrum, the explanation 'your submitting an application caused you to get a rejection letter' will be of no interest because it provides too little information. As we shall see in Section \ref{Practice}, getting clear on the relation between transparency and autonomy can help us get clear on the degree of transparency required in a given situation, as well as on how to compare the relevance of different causal explanations.

One final point before continuing. Readers familiar with the XAI literature might classify our definitions as focusing on so-called `\emph{post hoc} explanations'. This label would be somewhat misleading. Causal explanation can be available \emph{before} the fact in hypothetical form, and events that never occurred can be causally explained in a similar way. For example, we can say `if candidate A were to apply this year, she would not get the job because she has not finished her studies', thus providing a hypothetical causal explanation for an event that may or may not occur in the future. We are often interested in exactly such hypothetical causal explanations. A future applicant is well-advised to ask `if I were to apply, which factors will play a causal role in the outcome of my application?'. As will become apparent in the examples to follow, the right degree of transparency will often require that we grasp such hypothetical causal explanations. Our account of transparency focuses on explainability, but this does not restrict us to explainability after the fact.

\subsection{Causal explanation and autonomy}

Starting from our account, the demand for transparency translates into a demand for causal explanations. We can now ask what the value of causal explanation is: why do we or should we want causal explanations of AI algorithm outcomes? Recent work on causal explanation points toward an intuitive answer. Philosophical and empirical research has converged on the thesis that we are particularly interested in causal explanations because they provide us with reliable means to affect and predict our surroundings \citep[cf.][]{Lombrozo2011, Hitchcock2012}. For example, knowing that my subpar references caused me to not get the job allows me to predict that similar jobs will be unavailable to me unless I fix my references. It also provides me with an effective strategy to improve my chances of getting a similar job: improving my references. In effect, opaque decision algorithms hide effective strategies of affecting and predicting their outcomes from the affected parties. Conversely, transparent decision algorithms can enable us to undertake action and affect future outcomes.


In the XAI literature, this action-enabling potential of transparent decision algorithms often goes unmentioned. For example, a recent meta-study of over 100 texts on XAI by \citet{Langer2021}, lists twenty-eight desiderata found in the literature but makes no mention of how opacity can undermine the ability of affected parties to effectively influence the outcomes according to their goals.\footnote{They do list several texts that focus on the autonomy of those who deploy the algorithms, but none that raise a similar worry for the affected parties \citet[p.~7]{Langer2021}. As we shall see in Section \ref{Practice}, this focus on users rather than affected parties reverberates in the European Commission's AI act.} Similar remarks apply to a recent review of thirty-four AI ethics documents published by civil society, the private sector, governments, intergovernmental organizations, and multi-stakeholder organizations by \citep{Fjeld2020}.\footnote{The results of these meta-studies are in tension with Selbst and Barocas \citeyearpar{Selbst2018} claim  that transparency's action-enabling potential `dominates' the XAI literature (p.~1120). They refer to \citet{Wachter2017a}, who indeed dedicate a section to this topic, but further evidence of such dominance is scant.} \citet{Wachter2017a} is an exception to this pattern and contains a section on transparency as enabling users to alter outcomes. However, Wachter et al. do not explain why our ability to affect such outcomes is important.

This matter deserves closer attention. Our ability to reliably affect and predict the outcomes of decision algorithms can carry substantial moral weight. When it comes to pivotal decisions, such as whether I get a certain job, whether I am allowed into certain college, or whether I get an enormous insurance premium, my not being privy to what factors into these decisions undermines my ability to effectively shape my life. As rational planning agents, we attempt to influence such weighty decisions by preparing accordingly. We study to get the right degree, we select extra-curricular activities that are taken to sharpen the relevant skills, we drive carefully, and so on.\footnote{Though one would hope that we drive carefully for other reasons as well.} Our ability to shape our lives according to our plans and desires and to pursue our goals through deliberate actions, i.e. our ability to be  \emph{autonomous} agents, is severely undermined when we are denied information about what factors into life-changing decisions. There are at least three reasons for taking opacity's threat to our autonomy seriously. First, resolving the other worries concerning transparency does not automatically resolve the autonomy worry. Second, the autonomy worry comes with significant backing from moral philosophy. Third, undermining autonomy undermines responsibility. Let us discuss these points in turn.

First of all, the autonomy worry hones in on a feature that is tightly related to the opacity of decision algorithms. In principle, an opaque decision algorithm could be sufficiently reliable, fair, and trusted. Trust can be due to the recommendation of a trusted authority, and reliability and debiasing are eventually just a question of tweaking the algorithm. Certainly, tweaking the algorithm will be harder when we don't have the slightest idea how it works, and in any real-life scenario the programmers would require a minimum of transparency to get going, but there is nothing that \emph{in principle} stands in the way of an algorithm fulfilling its function perfectly without the relevant stakeholders having knowledge of what produces its outcomes. The outcomes could even be treated as negotiable by allowing users to double-check the outcomes using another decision algorithm, be it a human or an artificial one.\footnote{\citet[p.~98]{Wachter2017} make a similar suggestion to address cases where the demand for transparency conflicts with trade secrets.} Such double-checking can also be used as a basis for holding those who deployed or developed the algorithm accountable. By contrast, our lack of knowledge of how to \emph{affect} the outcomes is integral to an AI system being opaque to us. According to the causal account proposed above, opacity and a lack of knowledge about how to manipulate are definitionally inseparable. 


The upshot is that even if an opaque algorithm manages to tick all the other boxes, it can still threaten our autonomy. Suppose for example that all applications for government jobs go through the GOV-1 decision algorithm. For the purpose of our example, it matters little how GOV-1 was trained, but we can suppose that GOV-1 is the most reliable system to select government job candidates. All GOV-1 requires to decide who is the right candidate for the job is an accurately filled out questionnaire for each applicant. Careful analysis has demonstrated beyond reasonable doubt that no team of humans or competing artificial algorithms (or any combination thereof) will select a more viable candidate for any government position than GOV-1. We can assume further that applicants have a right of redress by demanding their applications be considered by another (non-AI) system, the government is held accountable for any mistakes by GOV-1 and we can assume trust in the government is strong enough to engender trust in GOV-1 as well. Unfortunately, it is unclear how GOV-1 weighs the information provided in the questionnaire. As a prospective applicant, I have no idea which competences I should acquire in order to become desirable, or even eligible, for a government job.\footnote{See \citet[Ch. 6]{ONeill2016}, \citet{Van2019} and \citet{Heilweil2019} for discussions of actual AI recruitment systems such as Hirevue, ZipRecruiter and Kronos.}

The opacity of GOV-1 thus hides salient ways of shaping our lives. Perhaps my goal of becoming a government employee requires me to focus more on getting into an international exchange program than on getting high grades. Perhaps I should not have given up my role in the student union to take an extra math credit. These are relatively simple examples, but AI systems can pick up on patterns that are very hard for us to detect. Consequently, GOV-1 might value features that one would assume to be completely irrelevant for being a good candidate. In so far as GOV-1 is opaque to me, I have no way of knowing, and no way of properly weighing my professional goals against any other goals I might have in life: building a family, maintaining friendships, learning to play the sitar\ldots{}  More generally, withholding explanations of decisions amounts to withholding information about how to affect these decisions. When opaque algorithms are used in life-changing decisions, they thus obstruct us from effectively shaping our lives according to our preferences. In short, opaque algorithms can undermine our autonomy, even when they respect other requirements such as reliability, fairness, accountability, and negotiability.

A second reason for focusing on opacity's threat to autonomy is that the moral import of autonomy is well-discussed in the philosophical literature. Autonomy takes center stage in several ethical theories. In areas varying from foundational deontology \citep[e.g.][]{Kant1993} and utilitarianism \citep[e.g.][]{Mill1999}, to more applied work on education \citep[e.g.][]{Haji2008, Thorburn2014}, bioethics \citep[e.g.][]{MacKay2016}, political theory \citep[e.g.][]{Raz1986}, and free will \citep[e.g.][]{Ismael2016} authors agree that autonomy carries moral weight.  The importance of autonomy is picked up in other discussions within AI ethics \citep[e.g.][]{Kim2021}, and legal theory as well \citep[e.g.,][]{Marshall2008, Mclean2009}. The call for transparency can draw support from all of these fields.

This appreciation of autonomy's moral importance gave rise to many accounts of what it precisely consists in. It is a further question whether our use of `autonomy' here overlaps or coincides with how it is standardly used in the literature. A detailed comparison will unfortunately have to wait for some other occasion, but here are two remarks on the subject. First, autonomy is taken to be some brand of self-determination and there is growing attention for the fact that self-determination requires a long-term control over one's life and plans \citep[e.g.][]{MacIntyre1983,Raz1986, Atkins2000, Bratman2000, Bratman2018, Ismael2016}. Second, even authors whose accounts would not appear to mesh well with our usage here acknowledge that cases like GOV-1 need to be taken into account. For example, \citet{Christman1991} defends a broadly internalist view of autonomy, according to which our desires should fit our rational beliefs and the rationality of a belief does not require its being true. So in principle, prospective applicants can rationally believe that studying political science rather than physics will help their chances, even if it does not. Even so, Christman accepts that rationality requires a somewhat reliable connection to reality:
\begin{quote}
One is autonomous if one comes to have one's desires and beliefs in a manner which one accepts. If one desires a state of affairs by virtue of a belief which is not only false but is the result of distorted information given to one by some conniving manipulator, one is not autonomous just in case one views such conditions of belief formation as unacceptable (subject to the other conditions I discuss) \citep[p.~16]{Christman1991}.
\end{quote}
Generally speaking, it would be surprising if no lack of information could undermine our autonomy.  Based on these two observations, one can reasonably expect that autonomy as the ability to shape one's life will correspond sufficiently with common usage in ethics.


A third reason to focus on the threat to autonomy is that autonomy strongly correlates with responsibility. Generally speaking, undermining an agent's autonomy relative to an outcome undermines responsibility relative to that outcome as well. This is because responsibility for an outcome requires a reliable causal connection between the agent's intention to reach or avoid the outcome and the outcome in fact being reached or avoided \citep[cf.][]{Bjornsson2012, Bjornsson2013, Grinfeld2020, Usher2020}. If candidate A intends to get a government job, but has no idea how they should polish their competences in order to qualify, the reliability of the correlation between her intending so and her achieving her goal should be expected to decrease. Generally speaking, not knowing how to achieve a goal makes it less likely that you achieve it. When opaque algorithms undermine our autonomy, they also undermine our responsibility for the outcome.\footnote{See \citet{Baum2022} for a worked out argument along these lines with a focus on users in the loop rather than affected parties.}

In conclusion, it appears that opacity can undermine autonomy and autonomy has moral value. Even if we set aside the above-mentioned connection with trust, fairness, reliability, accountability, negotiability, and a primitive preference for explanation, demands for transparency can still be grounded in a requirement to respect personal autonomy.

\section{Transparency in practice}\label{Practice}
Grounding the demand for transparency in autonomy requirements sheds new light on some familiar issues in XAI. I elaborate on three of these here: (i) the question how much transparency is desirable and whether this degree is technically attainable, (ii) how to entrench the demand for transparency legally, and (iii) whether transparency conflicts with other desiderata decision procedures. Let us discuss these in turn.

\subsection{Delivering the right degree of transparency}
The connection between transparency and autonomy provides guidance in deciding how transparent decision algorithms ought to be. We want to know how differences in the input correlate with differences in the output. It is well-received in both the philosophical and the computer science literature that knowledge of such higher-level regularities does not presuppose knowledge of the finer details of the system \citep[e.g.][]{Dennett1971, Dennett1991, Newell1982, Campbell2008}. Establishing which level of explanation is appropriate for the relevant input-output correlation will no doubt be an arduous task that requires different strategies for different cases, but there is a general point to be made here as well. In order to increase or maintain our autonomy, we want to know which changes in the input robustly correlate with certain changes in the output. Some patterns of correlation will be too fragile to be of genuine interest. Perhaps having a twitter handle without numerals increases your chances with 0.02 percent if you are a Caucasian woman with a law degree from a foreign university, but will have no effects in any other circumstances. Other patterns will be crucial knowledge for future applicants. Perhaps the only way to get a government job without a university degree is when you score above 150 on the IQ test and can demonstrate a staunch unwillingness to believe conspiracy theories. That is to say, in most circumstances, a university degree is a condition \emph{sine qua non}. Such robust correlations are worth knowing.

It is unlikely that knowledge of such robust correlations will require full physical details or design details, but it is also unlikely that the required explanation will always be found at the intentional level. This is not only because AI systems might typically lack such a level altogether (cf. supra), but also because the most robust patterns might not be found at this level. To see this, consider the human case of implicit bias. While such biases are not typically implemented at the intentional level, they can still make for robust correlations between certain input features and outputs. For example, committees with racist implicit biases might systematically review candidates with foreign-sounding names unfavourably, without those biases being manifested at the intentional level \citep[cf.][]{Bertrand2004}. The upshot is that there is no fixed level that provides the right level of transparency. Instead, we should focus on finding those correlations that are robust across the scenarios in which the algorithm is to be applied. 

The good news is that such input-output difference-making transparency is easier to achieve than full physical, design or algorithmic transparency. Full physical transparency would require a grasp of the workings of the system in all its physical details down to the electrons making up the hardware. Achieving such transparency is of course very difficult, but also not very useful. Full algorithmic transparency requires a grasp on the mathematical details of the algorithms encoded in the AI system and design transparency requires an engineering perspective on how such a system can be developed. Attaining either of these is taken to be extraordinarily difficult as well, and demanding it might even be in tension with copyright laws \citep[cf.][]{Burrell2016, Ananny2018, Wachter2017a}. By contrast, strategies for attaining transparency in input-output correlations are available. For example, \citet{Tubella2019} propose to test AI system's conformity with moral norms by using a `glass-boxing' method to check whether the input-output correlations of AI algorithms conforms to moral norms. A glass box is built around a system by checking its inputs and outputs. In Tubella et al.'s proposal, this glass box should detect input-output pairs that violate moral norms, but attaining input-ouput transparency requires less work. All the glass box should do is report the correlations between differences in inputs and differences in output. As the glass box is `built around' the AI system, this method would not require us to `open up' the algorithm that is being tested. Standard causal extraction algorithms, such as those developed by \citet{Pearl2000} and \citet{Spirtes2000}, can be used to acquire causal information on the basis of the correlational data gathered via glass-boxing. As \citet{Woodward2003} bases his account of causal explanation on the structural equation models provided by \citet{Pearl2000} and \citet{Spirtes2000}, the causal account of transparency and opacity we proposed in Section \ref{Transparency} naturally fits this technical approach.\footnote{Admittedly, such extraction algorithms require a certain causal structure to the data set in order to be successful. This need not be a problem as we typically have access to such minimal information. We know, for example, that for each individual case, the inputs cause the outputs, and not \emph{vice versa}.} Several other strategies that do not require `opening up' the black box have been developed \citep[cf.][]{Guidotti2018, Ustun2019, Belle2021}. Insofar as these strategies can in fact provide us with robust patterns of correlations between inputs and outputs, they can help safeguard user autonomy.\footnote{This is not to say that other concerns with AI opacity, such as assessing reliability and fairness will not require us to open the black box. As I have argued above, these are separate issues }

Undivided optimism about such `forensic' approaches would be premature. First of all, these approaches all omit details about the actual process leading up to the outcome when providing potential explanations. Such neglect of detail is necessary to provide explanations that are understandable for human agents, and, so the worry goes, there is a real risk that the omitted details are in fact crucial to the true causal story of how the decision was in fact produced \citep[cf.,][]{Rudin2019}. In the worst case, this lack of detail may make for mistaken explanations altogether, taking mere correlation for causation. While this is a real risk, it is worth noting that this risk is by no means unique to complex algorithms. It is well-recognized that the explanations of any event will require us to omit enormous amounts of details that, strictly speaking, contributed to its coming about \citep[e.g.][]{Loewer2007, Ney2009}. There is always a risk of omitting crucial details and focusing on irrelevancies. If we are to dismiss forensic approaches in general just because they are at risk of getting things wrong, we would be holding AI decision algorithms to a far higher transparency standard than we hold any other kind of explanation. The relevant question is whether these algorithms can reliably provide the right explanations in the contexts where they are implemented. Whether this can be done is an outstanding empirical question. If we are to use AI to make decisions that significantly impact our lives, this question needs to be investigated in-depth on a case-by-case basis.

\cite{Rudin2019} formulates a more distinctive challenge for current forensic approaches as well. Their ability to provide counterfactual information about what would have changed the outcomes of a particular procedure often relies on the notion of a `minimal' difference to the input \citep[e.g.,][]{Ustun2019}. The underlying idea is that users and affected parties are mostly interested in how they could have changed the outcome with as little effort as possible: I'd rather drastically improve my chances at getting a job by taking an evening course in business English than by taking a full law degree, for example. However, what counts as a `minimal' difference for the affected party is highly dependent on their context. Perhaps one's family context makes it easier to get a higher-paying job than to move to another ZIP code in order to get a mortgage. It thus probably won't do to provide a one-size-fits-all explanation for each outcome that just mentions the one difference-maker the algorithm deems `minimal'. A wider variety of realistic strategies to affect the outcome should be made available, such that the affected party can select the strategy that fits them best, if any. Whether providing such a smörgåsbord of options is practically feasible is, again, an open question.

There are ongoing attempts to make provide explanations of black box outcomes without `opening' black boxes. In order for these `forensic' strategies to safeguard autonomy, they need to reliably provide accurate explanations that fit the needs and practical perspective of the affected parties.




\subsection{Legally entrenching transparency demands}
Transparency demands play a central role in recent attempts to regulate the use of AI. However, the current formulations of these demands will not suffice to respect the autonomy of affected parties, as they do not provide the them with a right to explanation. Focusing on the European Union General Data Protection Regulation (GDPR) \citep{EUAI} and the European Comission's AI act \citep{GDPR}, \nocite{Annexes} I briefly elaborate on some of the obstacles for legally entrenching the relevant transparency demands. There is some hope that its link with autonomy can provide the call for transparency with extra legal backing, but there is still a long way to go.

There are two salient challenges with legally entrenching a right to explanation that will sufficiently respect the autonomy of the affected parties. First of all, if the AI algorithms in use are under protection of proprietary laws, the developers and deployers can maintain that any demand to disclose the workings of the algorithms conflicts with their right to intellectual property. True, the `forensic' tools discussed above could allow for the relevant causal knowledge without detailed knowledge of the algorithmic implementations. But making the case that the relevant causal knowledge can be acquired without the detailed knowledge that is defended by proprietary laws is likely to be an uphill battle. If for each demand for an explanation it has to be shown that the required explanation would not be in conflict with trade secrets, the affected parties are likely to find themselves \emph{de facto} disenfranchised with regards to their right to explanation. \citet[p.~98]{Wachter2017} make note of this challenge and suggest that external auditing mechanisms can be employed in such cases, but while such mechanisms could help to enforce the right of redress and to justify accusations of bias, they would not provide explanations, and thus not provide affected parties with action-enabling information.

Second, our previous discussion on finding the right level of explanation further complicates the picture. Any outcome will have many explanations that are of no practical use to the affected parties. For example, one causal explanation of why I did not get a government job might be that my application activated node 64879508, which activated node 0540324875, but inhibited node 45009783245. And, as it happens, such an activation pattern provides a negative outcome. Without any background knowledge about the role of these nodes within the system, this explanation is of no help to me, even if it is factual. The upshot is that the right to \emph{an} explanation will not suffice. The explanation needs to fit the explanatory needs and perspective of the affected party. While there are several excellent guides on how to provide explanations that are fitting in this sense \citep[e.g.,][]{ICO2021}, it remains an open question whether demands for a `fitting' translation can be legally entrenched. Even if a right to \emph{an} explanation can be made to go through, this further dimension needs to be kept in mind.

These difficulties with entrenching an adequate right to explanation transpire in recent attempts to regulate AI use. \cite{Wachter2017} point out that the GDPR fails to provide affected parties with a right to explanation of individual decisions, but instead delivers a rather vague right to be informed of (i) whether an AI is used, and (ii) what `logic' underlies the algorithm. Moreover, even this watered down right appears to apply only when the decision is based \emph{solely} on automated algorithms (Art.22(1)). Which means that even a minimal involvement of a human agent in the process can relieve the users of any obligation to provide information about the algorithm \citep[cf.,][p.~78]{Wachter2017}. I refer the reader to the original text for a detailed discussion of their evidence. Suffice it to say that the kinds of explanations that can sustain our ability to effectively pursue our life plans would not be protected by the GDPR alone.

More recently, the European Commission's AI act proposes to impose transparency requirements on precisely those algorithms with outcomes that have significant impacts on the affected parties, such as decisions on legal status, access to education or contractual relations (Annex III).\footnote{Note though, that the demands on AI decisions in `non-high risk' contexts are limited to mere guidelines that are adopted on a voluntary basis. This means that affected parties must be able to convincingly argue that the decisions affecting them count as `high risk'. This risks putting an undue burden of proof on victims who do not have the means to litigate in grey area cases.} However, as argued by \cite{Fink2021}, the proposal, if accepted, would only demand transparency towards the \emph{users} of AI, and not towards the affected parties.\footnote{This is despite the fact that the document explicitly distinguishes between users and affected persons (p.~30). This oversight is reminiscent of the general oversight in the texts reviewed by \citet{Langer2021}, which focused on the autonomy of the users, but made no mention of the autonomy of the affected parties (see Section \ref{Transparency}).} The only obligation towards the affected parties would be to inform them of the fact that an AI decision algorithm has been used to reach a conclusion. The AI act thus does little to legally entrench our autonomy in the face of decision algorithms that are entirely opaque to us.

There is some hope that a right to an adequate explanation of automatized decisions can be derived from more general rights. \cite{Fink2021} points out that article 41.(2c) of the European Charter of Fundamental Rights concerns exactly the obligation of the administration to ``give reasons for its decisions''. However, this article focuses on a right to good \emph{administration}, where `administration' is taken to refer ``the institutions, bodies, offices and agencies of the Union.'' Therefore, it is of no help to those who seek reasons for decisions made by insurance companies, banks, or private educational facilities. Moreover, the articles in the charter are explicitly restricted to cases where the administration are ``implementing Union Law'' (Art.51), which puts a further burden on affected parties to argue that the governmental decision they demand explained counts as an implementation of Union Law. There are several legal cases where a plaintiff's reliance on the right to good governance as spelled out in Art.41 is deemed illegitimate because the governmental decisions in question were not considered to count as implementations of Union Law.\footnote{See for example \cite[Sect.~17]{RvS2017}. Similar arguments are found in the Belgian Council of State decisions on cases 232.758 (29.10.2015), 233.512 (19.01.2016), and 238.292 (23.05.2017).} So, even when it concerns governmental decisions, Art.41(2c) of the charter provides only a limited right to explanation.

In light of the precarious position of the right to explanation in current legislation, the link between transparency and autonomy can perhaps be of help. There are at least some promising signs to be found here, as the notion of personal autonomy plays a central role in a variety of legal frameworks. For example, the European Court for Human Rights has relied on the notion of personal autonomy in several rulings,\footnote{See \citet{Koffeman2010} for an overview} and some legal scholars maintain that the very right to personal autonomy is enshrined in the European Convention of Human Rights (ECHR) \citep[e.g.][]{Marshall2008} and the US Constitution \citep[e.g.,][]{Mclean2009}. If respecting personal autonomy requires providing adequate explanations for potentially life-changing decisions, then the right to personal autonomy requires a right to explanation.

However, there are significant obstacles on this route towards a right to explanation as well. Neither the US Constitution or the ECHR explicitly mention autonomy. Instead, arguments for the right to autonomy based on these texts tend to go via related notions, such as dignity (ECHR, Art.1) and privacy (ECHR, Art.2). Consequently, the jurisprudence at the ECHR is equivocal on how much weight is to be attached to autonomy. Some judges rely on autonomy as a guiding notion to interpret these foundational legal texts, whereas others take the right to autonomy to follow from texts themselves.\footnote{See  \citet[p.~8--9]{Koffeman2010} for discussion.} It certainly would not hurt to have more concrete formulations of the importance of autonomy and a legal right to explanation that is adapted to the current rise of automated decision-making available in our legal frameworks.

While the call for transparency has inspired the notion to play a central role in legal documents focusing on AI, these documents do not guarantee transparency towards affected parties whose lives are affected by the outcomes of automatized decision procedures. The link between transparency and autonomy provides a promising extra tool for legally entrenching a right to explanation. Even so, there is still plenty of legislative work to be done before we can feel safe that such a right is legally protected.

\subsection{Downsides of transparency}
Even if transparency has a distinct moral value in serving autonomy, it should not be pursued at all costs. There may be any number of reasons to trade off transparency for other goods. I focus here on three potential drawbacks of providing the degree of transparency that is required to bolster autonomy.



First, providing information of the robust correlations that allow us to affect the outcomes of decision algorithms might increase the advantage of those who have easier access to the difference-making features. For example, if attending expensive private universities increases one's chances of getting a government job, it might be overall justifiable to not divulge this fact. As it becomes easier to control the outcomes in fair ways, it will become easier to control the system in unfair ways as well. Implementing transparency requirements will require consideration of this fact \citep[cf.][Sect.~3.2]{Weller2019}.

Second, it is a well-received truism that otherwise reliable indicators can become unreliable once it is publicly known that they are used as indicators \citep{Campbell1979, Goodhart1984}. For example, if word gets out that GOV-1 treats participation in debate club as a big plus, this might cause students to attend debate club merely to improve their chances at a government job, rather than to develop the relevant skills that debate club is supposed to foster, like absorbing and structuring information, and presenting clear arguments. The influx of new students with less intrinsic motivations to acquire these skills might drastically decrease the reliability with which debate clubs cultivate these skills in their participants, hence making participation in debate club a less reliable indicator. Making criteria used in life-changing decisions accessible to affected parties might similarly change the reliability of those same criteria.

Third, building on work by \citet{ONeill2002}, Nguyen (forthcoming) \nocite{NguyenForthcoming} recently argued that transparency can improperly limit the kinds of reasons that feature in decision-making. Forcing experts to make their reasons for certain judgments accessible and understandable for non-experts, risks forcing them to limit their reasons to the kind of reasons that non-experts are sensitive to. This would in effect make it impossible to invoke reasons that require advanced expertise to appreciate. While Nguyen does not focus on automatized decision-making, his arguments appear to transfer quite easily to the automatized case. In fact, one of the oft-cited reasons for using automated decision algorithms is that they appear to discover patterns that are not readily appreciable by human observers. If we only use transparent algorithms, we might come to see that some unexpected or simply intractable reasons feature in the production of the decision. In the long run, this might incite us to use less reliable mechanisms that only employ reasons and inferences that are tractable to us.

So even though transparency can help bolster autonomy, this does not mean that it is to be pursued at all costs. Relatedly, it is worth emphasizing that the argument presented here does \emph{not} establish that transparency or autonomy are \emph{intrinsically} valuable. The argument I have provided in favour of transparency is clearly restricted to its instrumental value. I have argued that transparency carries moral weight as a means for supporting user autonomy. The argument leaves it open whether or not (i) transparency is intrinsically valuable, (ii) autonomy is intrinsically valuable, or (iii) transparency has  instrumental value beyond its contribution to autonomy. \citet{Colaner2021} recently argued in favour of (i), and many of the works referenced throughout this text provide evidence that (ii) and (iii) are true as well. Even so, our central argument goes through even if (i)--(iii) turns out to be false: opaque decision algorithms can undermine our autonomy by hiding salient pathways of affecting their outcomes. If the broad consensus that autonomy carries moral weight is correct, this means transparency is worth demanding.

\section{Conclusion}
There are several reasons to demand that impactful decision algorithms are transparent. This is true regardless of whether they are implemented by humans or AI systems. Previous research indicated that transparency is conducive to negotiability and accountability, helps fine-tune reliability and avoid bias, and that we are more likely to trust the results of algorithms if we understand what causes their outcomes. Building on recent work about the value of explanation, I have argued that a lack of transparency can also undermine the autonomy of the affected parties. In both the human and the AI case, we are interested in knowing the robust patterns of correlation that allow us to reliably affect and predict their outcomes. Such knowledge can play an integral part in planning and shaping our lives as rational, self-determining agents.

This perspective on transparency furnishes XAI debates with both new tools and new challenges. On the one hand, calls for transparency can draw on established work in moral philosophy and legal texts that emphasizes the importance of personal autonomy. Moreover, focusing on the autonomy of affected parties can guide us in deciding what kinds of explanations we should demand. On the other hand, it appears that resolving previous concerns about opacity will not automatically address the threat to the autonomy of the affected parties, and the kinds of explanations required to respect our autonomy can be hard to come by. It might not require us to open the black box, but it does require us to take into account the perspectives of rational planning agents with life-goals and dreams.

\end{document}